\documentclass[twocolumn,aps,pra]{revtex4}
\usepackage{epsfig}
\usepackage[english]{babel}
\usepackage{latexsym}
\usepackage{graphics}
\usepackage{subfigure}
\usepackage{graphicx}
\usepackage{dcolumn}
\usepackage{amsmath}
\usepackage{hyperref}
\usepackage{amssymb}
\usepackage{color}

\begin{document}

\title{Alignment dependence of photoelectron momentum distributions for diatomic molecules N$_2$ in strong elliptical laser fields}

\author{Dianxiang Ren$^{1,\S}$, Shang Wang$^{2,3,\S}$, Chao Chen$^{2}$, Xiaokai Li$^{1}$, Xitao Yu$^{1}$, Xinning Zhao$^{1}$, Pan Ma$^{1}$,
Chuncheng Wang$^{1}$, Sizuo Luo$^{1,*}$, Yanjun Chen$^{2,\dag}$, and Dajun Ding$^{1,\ddag}$}

\affiliation{1.Institute of Atomic and Molecular Physics, Jilin University, Changchun, China\\
2.College of Physics and Information Technology, Shaan'xi Normal University, Xi'an, China\\
3.College of Physics and Hebei Key Laboratory of Photophysics Research and Application, Hebei Normal University, Shijiazhuang, China}

\date{\today}

\begin{abstract}
We study ionization dynamics of aligned diatomic molecules N$_2$ in strong elliptical laser fields experimentally and theoretically.
The alignment dependence of photoelectron momentum distributions (PMDs) of N$_2$ measured in experiments is highlighted with comparing to Ar measured synchronously.
Our results show that the PMDs of N$_2$  depend strongly
on the alignment of the molecule, relative to the main axis of the laser ellipse. In particular,
the most-probable electron-emission angle which is often used in attosecond measurement,
differs remarkably when changing the molecular alignment.
We show that the interplay of two-center interference and tunneling when the electron goes through the laser-Coulomb-formed barrier,
plays an important role in these phenomena.
Our work gives suggestions on studying ultrafast electron motion inside aligned molecules.

\end{abstract}

\maketitle

\section{Introduction}
Tunneling is one of the most fundamental physical processes in the interaction of  atoms and molecules with strong laser fields.
The electronic wave packet created through tunneling triggers rich ultrafast dynamical processes such as  above-threshold ionization (ATI)
\cite{above0,above1,above2,above3}, autoionization \cite{auto1,auto2}, photoelectron holography \cite{Holography1,Holography2,Holography3,Holography4},
photoelectron diffraction \cite{diffraction1,diffraction2,diffraction3}, nonsequential double ionization (NSDI) \cite{DOUBLE1,DOUBLE2,DOUBLE3,DOUBLE4,DOUBLE5},
and high-harmonic generation (HHG) \cite{McPherson1987, Huillier1991,Corkum,Lewenstein1994}, etc. Using these processes, people can probe the structure of
the target and trace the electron motion in atoms and molecules on the attosecond time scale, opening the new research region in ultrafast science,
namely attosecond physics \cite{Krausz2009,Krausz,Vrakking}.
Therefore, studies on the electron dynamics inside atoms and molecules in the tunneling process are of basic significance
for understandings and applications of these tunneling-triggered strong-field processes.

Strong-field tunneling can be well described with theories of ADK \cite{ADK1,ADK2}, PPT \cite{PPT} and strong-field approximation (SFA) \cite{Lewenstein1995,Becker2002}.
These theories provide deep insights into roles of structure of the target in tunneling ionization.
In the frame of SFA with saddle-point theory, the motion of the tunneling electron under the barrier, which is formed by the laser field and the Coulomb potential, is characterized by an imaginary momentum.
For aligned symmetric diatomic molecules \cite{Alignment-Dependent1} with a not very large internuclear distance $R$, it has been shown that
 the interference of the electronic wave  between these two atomic centers of the molecule under the barrier
induces the remarkable increase of ionization yields of the target molecule in comparison with its partner atom with a similar ionization potential $I_p$ \cite{Chen2009,Chen2010,Chen2012}.
Further theoretical studies showed that this effect  has important influences on alignment dependence of ATI \cite{Gao2017,Ren2019}, HHG \cite{Chen2018} and NSDI \cite{Li2012} of symmetric molecules,
as well as ATI and HHG of oriented asymmetric molecules \cite{Xie2018}.

On the other hand, photoelectron momentum distributions (PMDs) in strong elliptically-polarized laser fields with large ellipticity have been widely used in probing the tunneling dynamics  of atoms and molecules,
and the relevant probing procedure has been termed as attoclock \cite{Eckle2008,Keller2008}.  In the attoclock experiments, the brightest part of PMD, which is associated
with the most-probable emission angle of the photoelectron, is generally used as the observable characteristic quantity through which the key dynamical information of the studied system is deduced \cite{Eckle2008,Keller2008,Pfeiffer,Torlina2015,Klaiber,Camus,Wang2017,Sainadh,Han,Serov2019,Khan2020,Yan2020,Che2021}. For example, with this characteristic quantity, the intriguing  issues  of  tunneling time \cite{Eckle2008,Keller2008,Torlina2015,Camus,Sainadh,Han,Serov2019,Khan2020,Yan2020}, tunneling exit \cite{Pfeiffer}, nonadiabatic effects in tunneling \cite{Klaiber},
excited tunneling \cite{Wang2017}, permanent-dipole effects in tunneling \cite{Che2021}, etc., have been explored deeply.

In the attoclock experiments of molecular targets, it is important to study the effect of two-center interference under the barrier on the most-probable emission angle
when the molecule is aligned along a specific angle relative to the laser polarization.
Since many dynamical properties of molecules in intense laser fields such as tunneling \cite{ADK1} and excitation \cite{Chen2012} depend strongly on the molecular alignment.
Such studies can be performed theoretically with assuming perfect alignment \cite{Ren2019}.
However, in experiments, perfect alignment is impossible and therefore a clear identification of the interference effect is not easy to achieve.

In this paper, we study ionization dynamics of the $\text{N}_2$ molecule in strong elliptical laser fields at different alignment angles $\theta$
($\theta$, the angle between the molecular axis and the major axis of the laser ellipse) experimentally, with synchronous measurements of N$_2$ and Ar.
Then we compare the experimental results with those obtained with  numerical solution of time-dependent Schr\"{o}dinger equation (TDSE) and a modified SFA model (MSFA)
which considers the effect of long-range Coulomb potential. 
To overcome the difficulty of imperfect alignment and highlight the effect of a specific alignment angle,
we analyze the relative PMD related to the difference between PMDs of aligned N$_2$ and Ar synchronously measured in experiments.
The relative PMD shows the strong dependence on the molecular alignment, and
the corresponding offset angle $\phi$ ($\phi$, the angle between the most-probable electron-emission direction and the minor axis of the laser ellipse)
shifts remarkably when the angle $\theta$ changes.
In addition, the distribution of transverse momentum of the molecule also differs remarkably for different angles $\theta$ and differs from that of the reference atom Ar.
These results are in good agreement with TDSE and MSFA predictions. Our further analyses show that the effect of two-center interference under the barrier
which is sensitive to the angle $\theta$ plays an important role in the alignment dependence of PMD of molecules, with remarkably affecting  the offset angle and the transverse momentum.

\section{Experimental and theoretical methods}

\subsection{Experimental setup}

Our experiment setup includes a pump-probe laser system and a cold target recoil ion momentum spectrometer (COLTRIMS)\cite{experimental1}.
The experimental device was introduced in our previous publications in detail \cite{auto2,experimental1,experimental2},
where the produced ions and electrons are coincidence measured.  The laser pulses in our experiments are generated from a Ti: sapphire femtosecond
laser system (${\sim}$35 fs) with a repetition rate of 1 kHz at the wavelength of 800 nm. The laser beams are focused on the supersonic beam
with a concave mirror (f=75 mm) to ionize the atoms and molecules. A linearly-polarized laser and an elliptically-polarized laser are chosen
as the pump and probe lasers. The pumping laser is stretching to around 300 fs by passing through a 5 cm BK7 glass. The polarization of probing laser
is achieved by adjusting the relative angle between half-wave plate and quarter-wave plate, and the ellipticity of the probing laser is set as $\varepsilon{\sim}0.8$.
The alignment of $\text{N}_{\text{2}}$ is triggered by the pumping laser and the revival of alignment is monitored by scanning the delay time between pumping and probing lasers \cite{experimental4,experimental6}.
The pump laser beam has a diameter of around 8 mm and the diameter of probing laser beam is 12 mm, this setup is to ensure that only aligned molecules
prepared by pumping laser are ionized by probing laser.
The brief process of experiment is exhibited in Fig. 1.  In order to compare ionization mechanism of molecule with atom,
a gas mixture containing $\text{N}_{\text{2}}$ and $\text{Ar}$ (2:1) is applied to reduce any possible difference of experimental conditions
in the measurements for diatomic molecules and their companion atoms. We keep the pumping laser to a sufficiently low intensity
so that ionization rate is low enough. The aligned samples ($\text{cos}^2\theta$ $\sim 0.65$) are prepared by setting the probing laser at $\sim$ 3.9 ps, 
the relative angle between the molecular axis and the major axis of ellipsoidal light is controlled by rotating the polarization of the pump laser.

\begin{figure}[t]
\begin{center}
\rotatebox{0}{\resizebox *{2.5in}{!} {\includegraphics {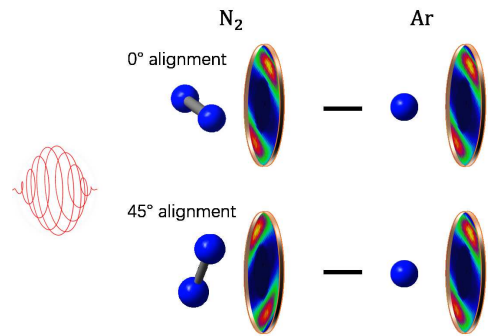}}}
\end{center}
\caption{Schematic diagram of  alignment-probe scheme for measuring alignment-dependent
ionization of ${\rm{N}}_2$ molecule.
} \label{Fig. 1}
\end{figure}

\subsection{Numerical methods}
The Hamiltonian of the  model molecule  studied here has the form of
\begin{equation}
H(t)=H_0+\text{\textbf{r}}\cdot \text{\textbf{E}}(t)
\end{equation}
with the field-free Hamiltonian $H_0=\textbf{p}^2/2+V(\textbf{r})$.
We use the soft-Coulomb  potential
\begin{equation}
V(\textbf{r})=-\sum\limits_{j=1,2}\frac{(Z_{ji}-Z_{jo})\text{exp}(-\rho{\textbf{r}^{2}_{j}})+Z_{jo}}{\sqrt{\xi +\textbf{r}^{2}_{j}}}
\end{equation}
with $\textbf{r}_{1,2}^{2}=(z{\pm}{\frac{{{R}}}{{2}}}{\text{cos}\theta})^{2}+(y{\pm}{\frac{{{R}}}{{2}}}{\text{sin}\theta})^{2}$ in two-dimensional cases. Here, $R$ is the internuclear distance.
The indices $i$ and $o$ denote the inner and outer limits of $Z_{1}$ and $Z_{2}$, $\rho$ is the screening parameter.
$\xi=0.5$ is the smoothing parameter which is used to avoid the Coulomb singularity. We have used the parameters \cite{Chen2018} of
$Z_{1i}$=$Z_{2i}=5$, $Z_{1o}=Z_{2o}$=0.5, $\rho=1.555$ and $R=2.079$ a.u. for ${\rm{N}}_2$ with the ionization potential of $I_p=$ 0.57 a.u..
For comparison, the parameters of $Z_{1i}=Z_{2i}=0.517$, $Z_{1o}=Z_{2o}=0$, $\rho=0$ and $R=0$ a.u. are also used for a model Ar atom with a similar $I_p$ to N$_2$.
 $\text{\textbf{E}}(t) = {E_{z} (t){\vec{\textbf{e}}}_z+E_{y} (t){\vec{\textbf{e}}}_y}$
with $E_{z}(t)=f(t)\frac{\varepsilon{_0}}{\sqrt{1+\varepsilon{^2}}}{\text{\text{cos}}\omega_0}t$ and $E_{y}(t)=\varepsilon{f(t)}\frac{\varepsilon{_0}}{\sqrt{1+\varepsilon{^2}}}{\text{sin}\omega_0}t$
is the laser electric field. $\varepsilon$ is the laser ellipticity and $\varepsilon{_0}$ is the laser amplitude relating to the peak laser intensity $I$.
$f(t)$ is the envelope function, ${{\omega }_{0}}$ is the laser frequency and ${{\vec{\textbf{e}}}_{z}}$(${{\vec{\textbf{e}}}_{y}}$) is the unit vector
along the major (minor) axis of the polarization ellipse. In our simulations, we employ a 15-cycles laser pulse which is switched on and off linearly
over three optical cycles and keeps a constant intensity for nine additional cycles.

We solve the TDSE of $i\dot{\Psi}(\mathbf{r},t)=H(t)\Psi(\mathbf{r},t)$ numerically through the spectral method \cite{spectral method}. The numerical details can be found in  \cite{Gao2017}.
In our simulations,  for ${\rm{N}}_2$, we select the eigenstate of $H_0$ with $3\sigma_g$ symmetry as the initial state, and for the model Ar atom,
the eigenstate of the corresponding $H_0$ with $1s$ symmetry is selected.
The ground state of the active electron of the real Ar atom has the $3p$ symmetry.
However, in our experiments, the Ar atom is randomly distributed, so it behaves similarly to an atom with $s$ symmetry.

\subsection{Strong-field approximation}
In the SFA, the amplitude of the photoelectron with the drift momentum $\textbf{p} $ can be written as \cite{Lewenstein1995}
\begin{equation}
c(\textbf{p})=-i\int^{T_{p}}_0dt^\prime{\textbf{E}}(t^\prime)\cdot{\textbf{d}_i}{[\textbf{p}+\textbf{A}(t^\prime)]}e^{iS(\textbf{p},t^\prime)}.
\end{equation}
Here, the term  $S(\textbf{p},t)=\int_{}^{t}\{{[\textbf{p}+\textbf{A}(t'})]^2/2+I_p\}dt'$ is the semiclassical action and $T_{p}$ is the length of the total pulse.
The term $\textbf{d}_i(\textbf{v})=\langle{\textbf{v}}|\textbf{r}\vert{{0}\rangle}$ denotes the dipole matrix element for the bound-free transition.
The term $\textbf{A}(t)=-\int^{t}\textbf{E}(t')dt'$ is the vector potential of the electric field $\textbf{E}(t)$.

For the laser field with high intensity and low frequency, the temporal integral in Eq. (3) can be evaluated by the saddle-point method \cite{Lewenstein1995,Becker2002},
with solving the following equation
\begin{equation}
[\textbf{p}+\textbf{A}(t_s)]^2/2+I_p=0.
\end{equation}
The solution $t_s$ of the above equation is complex which can be written as $t_s=t_{0}+it_{i}$. The real part $t_0$ can be understood as the ionization time,
and the imaginary part $t_i$ can be understood as the tunneling time.
The corresponding momentum-time pair ($\textbf{p},t_0$) have been termed
as electronic trajectory. The corresponding amplitude of the trajectory ($\textbf{p},t_0$) can be written as
\begin{equation}
F(\mathbf{p},t_0)\equiv F(\mathbf{p},t_s)\propto\big[\beta\textbf{E}(t_s)\cdot \textbf{d}_i(\textbf{p}+\textbf{A}(t_s))e^{iS}\big],
\end{equation}
with $S\equiv S(\textbf{p},t_s)$ and $\beta\equiv({1/det(t_s)})^{1/2}$.
The term $det(t_s)$ in the definition of $\beta$ is the determinant of the matrix formed by the second derivatives of the action \cite{Lewenstein1995}.
The whole amplitude for photoelectron with a momentum $\textbf{p}$ can be written as
\begin{equation}
c(\textbf{p})\propto \sum_s F(\textbf{p},t_s).
\end{equation}
The sum runs over all possible saddle points.

The SFA neglects the Coulomb effect which is important for the appearance of the nonzero offset angle in PMD of elliptical laser fields  \cite{Eckle2008,Keller2008}.
To consider the Coulomb effect, we use a MSFA model \cite{MSFA} following the semiclassical procedures in \cite{Brabec,Goreslavski,yantm2010}.
In this model, after tunneling, the electron motion in the combined laser and Coulomb fields is governed by the Newton equation
\begin{equation}
\ddot{\textbf{r}}(\textbf{p},t)=-\textbf{E}(t)-\nabla_{\textbf{r}}V(\textbf{r})
\end{equation}
for each SFA electron trajectory ($\textbf{p},t_0$) with initial conditions of exit momentum $\dot{\textbf{r}}(\textbf{p},t_0)=\textbf{p}+\textbf{A}({t_{0}})$ and exit position $\textbf{r}(\textbf{p},t_0)=Re\int_{t^{\prime}}^{t_{0}}[\textbf{p}+\textbf{A}({t^{\prime\prime})}]d{t^{\prime\prime}}$ at $t^\prime=t_{0}+it_{i}$.
The final Coulomb-modified  drift momentum is obtained with
\begin{equation}
\mathbf{p}_f=\dot{\mathbf{r}}(\mathbf{p},t\rightarrow\infty),
\end{equation}
with the amplitude $F(\mathbf{p}_f,t_0)\equiv F(\mathbf{p},t_0)$.
Accordingly, the whole amplitude for photoelectron with the Coulomb-corrected momentum $\textbf{p}_f$ in the present MSFA model can be written as
\begin{equation}
c(\textbf{p}_f)\propto \sum_s F(\textbf{p}_f,t_s).
\end{equation}
With considering the permanent dipole, the MSFA model can also be applied to polar molecules \cite{Wang2020}.

\subsection{Two-center interference in tunneling}
Next, we discuss the form of the dipole term $\textbf{d}_i(\textbf{k}_s)$ with $\textbf{k}_s=\textbf{p}+\textbf{A}(t_s)$ in Eq. (5) for the model N$_2$ molecule.
With the linear combination of atomic-orbitals-molecular-orbitals (LCAO-MO) approximations \cite{chen2009} and assuming that the molecular axis is along the  $z$ axis,
the ground state wave function of N$_2$ with $3\sigma_g$ symmetry can be written as
$\langle{\textbf{r}}{\arrowvert}0\rangle_{3\sigma_g}{\sim}{{z}_ae^{-\kappa{\text{r}_a}}}-{{z}_be^{-\kappa{\text{r}_b}}}$. Here, $\kappa=\sqrt{2I_{p}}$, ${z}_{a,b}=z\pm{{R}}/2$
and ${r}_{a,b}=|\textbf{r}\pm{\textbf{R}/2}|$. Then the dipole $\textbf{d}_i(\textbf{k}_s)$ for N$_2$, denoted with $\textbf{d}_{3\sigma_{g}}(\textbf{k}_{s})=\langle{\textbf{k}_s}|\textbf{r}\vert{{0}\rangle}_{3\sigma_{g}}$, can be written as \cite{Chen2018}
\begin{equation}
{\textbf{d}_{{{3\sigma_{g}}}}(\textbf{k}_{s})\propto{N_{3\sigma_{g}}}[-2i\text{sin}({\frac{\textbf{k}_{s}\cdot\textbf{R}}{2}})\textbf{d}_{2p_{z}}}(\textbf{k}_{s})].
\end{equation}
Here,   $\textbf{R}$ is the vector between these two nuclei of the molecule and ${N_{3\sigma_{g}}}$ is the normalization factor.
The term $\text{sin}({{\textbf{k}_{s}\cdot\textbf{R}}/{2}})$ describes the interference of the electronic wave between these two atomic centers of  the molecule,
and the term
\begin{equation}
\textbf{d}_{2p_{z}}(\textbf{k}_{s})=\langle{\textbf{k}_s}|\textbf{r}\vert{{0}\rangle}_{2p_{z}}
\end{equation}
is the atomic dipole moment from the $2p_{z}$ orbit.

Clearly, according to Eq. (4), for nonzero $I_p$, with the complex solution $t_s=t_{0}+it_{i}$, the momentum  $\textbf{k}_s=\textbf{p}+\textbf{A}(t_s)$ is complex,
which characterizes the tunneling process of the bound electron going through the laser-Coulomb-formed barrier.
For the elliptical laser field, the complex $z$ and $y$ components of the vector potential $\textbf{A}(t_s)$ can be written as ${A}_{z(y)}(t_s)={A}_{r,z(y)}(t_s)+i{A}_{i,z(y)}(t_s)$.
Then the complex momentum $\textbf{k}_s$ can be expressed as $\textbf{k}_s=\textbf{k}_s^r+i\textbf{k}_s^i$ with $\textbf{k}_s^r=[p_z+A_{r,z}(t_s)]\vec{\mathbf{e}}_z+[p_y+A_{r,y}(t_s)]\vec{\mathbf{e}}_y$
and $\textbf{k}_s^i=A_{i,z}(t_s)\vec{\mathbf{e}}_z+A_{i,y}(t_s)\vec{\mathbf{e}}_y$. Accordingly, the interference term  $\sin(\textbf{k}_{s}\cdot \textbf{R}/2)$
has the following form
\begin{equation}
\begin{aligned}
\sin(\textbf{k}_{s}\cdot \textbf{R}/2)=\sin(\textbf{k}_s^r\cdot \textbf{R}/2+i\textbf{k}_s^i\cdot \textbf{R}/2).
\end{aligned}
\end{equation}
With assuming $|p_z+A_{r,z}(t_s)|\sim0$ and $|p_y+A_{r,y}(t_s)|\sim0$, as in \cite{Gao2017}, we have $\textbf{k}_s\approx i\textbf{k}_s^i$.
By Eq. (4), we also have $|\textbf{k}_s^i|\approx\kappa$. Accordingly, we have
\begin{equation}
\begin{aligned}
\sin(\textbf{k}_{s}\cdot \textbf{R}/2)\approx\sin(\frac{i\kappa R \cos\theta_I}{2}).
\end{aligned}
\end{equation}
Here, $\theta_I$ is the angle between the vector
$\textbf{k}_s^i$ and the molecular axis. A similar expression to Eq. (13) have been used in \cite{Gao2017} to analyze effects of two-center interference in ATI in orthogonal two-color laser fields.
The approximate expression of Eq. (13) shows the interference of the electronic wave in tunneling, characterized by the imaginary momentum ``$i\textbf{k}_s^i$" with $|\textbf{k}_s^i|\approx\kappa$. The interference effect
depends on the molecular alignment as the angle $\theta_I$ is alignment dependent.
On the whole, the value of the term $|\sin(\frac{i\kappa R \cos\theta_I}{2})|=|e^{(\frac{i\kappa R \cos\theta_I}{2})}-e^{-(\frac{i\kappa R \cos\theta_I}{2})}|/2$ increases with the increase of the internuclear distance R
and the decrease of the angle $\theta_I$.
For a linearly polarized laser field, we have $\theta_I=\theta$, and the above expression clearly indicates  that the ionization is the strongest at $\theta=0^o$ and is near zero at $\theta=90^o$.
In the following, we will show
that the interference effect influences remarkably on the PMD in the elliptical laser field, especially for the offset angle.

\begin{figure}[t]
\begin{center}
\rotatebox{0}{\resizebox *{6cm}{5cm} {\includegraphics {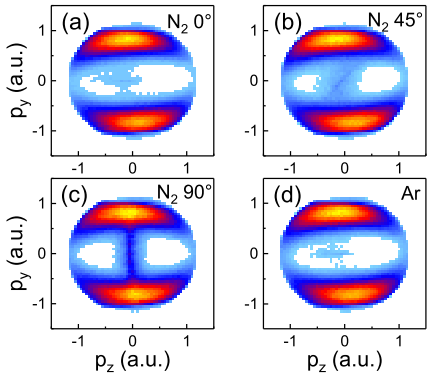}}}
\end{center}
\caption{
Normalized PMDs of aligned ${\rm{N}}_2$ measured at $\theta=0^{\text{o}}$ (a), $45^{\text{o}}$ (b) and $90^{\text{o}}$ (c).
PMDs of Ar are measured synchronously with N$_2$ at these angles. The case of Ar measured with N$_2$ at $\theta=0^o$ is shown in (d).
Other cases of Ar are similar to that shown here.
All of the results are normalized to the whole amplitude of the corresponding distribution.
}
\label{fig:2}
\end{figure}

Note, according to Eq. (5), the ATI from an atom with a $p$-style orbital such as $2p_z$ orbital, the wave function of which is not
spherically symmetric,   also depends on the angle between the laser polarization and the $z$ axis of the chosen coordinate system,
since the term ${\textbf{E}}(t_s)\cdot{\textbf{d}_{2p_z}}(\textbf{k}_s)$ is also angle dependent. However, in real experiments,
atoms can not be aligned and are randomly distributed, so the ionization of atoms is considered to be alignment independent.
The laser intensity used in our experiments is  $I{\sim}2\times {{10}^{14}}{\text{W}}/{\text{c}{{\text{m}}^{2}}}$.
In our theoretical calculations, we have used the laser parameters of
$I{=}2\times {{10}^{14}}{\text{W}}/{\text{c}{{\text{m}}^{2}}}$, $\lambda{=}800$ $\text{nm}$ and $\varepsilon{=}0.8$.

\begin{figure}[t]
\begin{center}
\rotatebox{0}{\resizebox *{8.5cm}{10cm} {\includegraphics {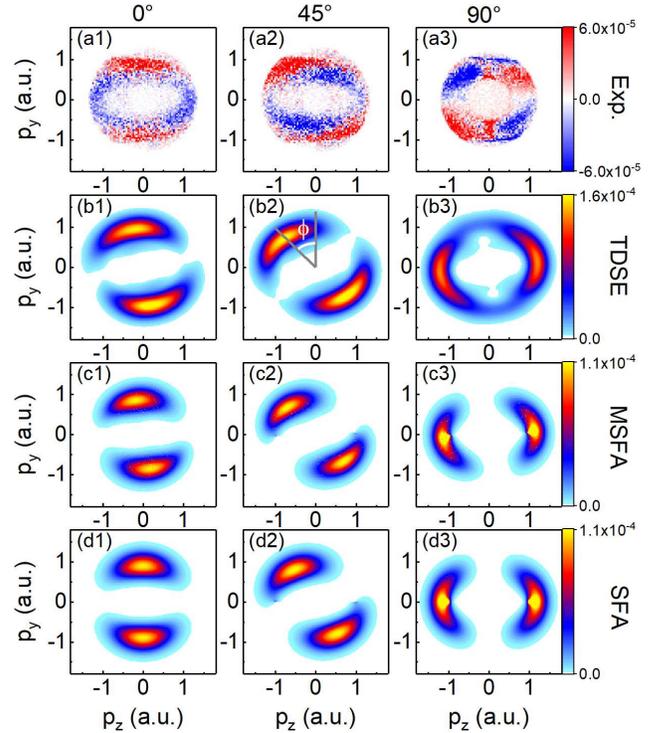}}}
\end{center}
\caption{
Comparisons of PMDs of N$_2$ at $\theta=0^{\text{o}}$ (the first column), $45^{\text{o}}$ (the second column) and $90^{\text{o}}$ (the third column),
obtained with experiments (the first row), TDSE (the second row), MSFA (the third row) and SFA (the fourth row).
Experimental results are the differences between normalized PMDs of aligned ${\rm{N}}_2$ and Ar measured synchronously.
Theoretical results are normalized to the whole amplitude of the corresponding distribution.
}
\label{fig:3}
\end{figure}

\section{RESULT AND DISCUSSIONS}

In our experimental studies for N$_2$, we focus on three typical alignment angles of $\theta=0^o$, $\theta=45^o$ and $\theta=90^o$.
However, the molecule can not be aligned fully in the pump step.
As a result, the measured PMD of N$_2$ at a certain alignment angle always contains the information of other angles,  
making the PMD of the aligned molecule  somewhat similar to the randomly-distributed atomic case.
To highlight the comparison for PMDs of N$_2$ at different angles, the PMDs of  N$_2$ aligned at a certain angle
and Ar are measured at the same time, as introduced in the method section. Then we evaluate the difference between the normalized PMDs of N$_2$ at the angle $\theta$ and Ar.

In Fig. 2, we plot the measured PMDs of  N$_2$ at these three typical angles and Ar.
In each panel, the alignment of the molecular sample relative to the main axis of the laser ellipse is also reflected at the center of the distribution  by a dark line.
All of the results are somewhat similar, with the PMDs showing a small counterclockwise rotation relative to the axis of $p_z=0$.
This rotation has been attributed to the Coulomb effect. For an atom, this rotation disappears in the general SFA simulations without considering the Coulomb effect.
However, the situation changes for the difference of aligned N$_2$ and Ar, as shown in the first row of Fig. 3.
In contrast to Fig. 2, the results differ remarkably from each other here.
When both of the results of $\theta=0^o$ and $45^o$ show an anticlockwise rotation structure for the bright part of the distribution,
with the brightest parts locating in the second and the fourth quadrants, the result of $90^o$ shows the clockwise one,
with the brightest parts in the first and the third quadrants.
In addition, the offset angle (which is defined as the angle between the axis of $p_z=0$ and the axis which goes through the origin and the brightest part of the PMD)
for the result at $\theta=45^o$ is larger than that at $\theta=0^o$. These remarkable differences between different angles are in good agreement with
the predictions of TDSE and MSFA, as shown in the second and the third rows of Fig. 3.
They are also reproduced by the SFA, as shown in the fourth row of Fig. 3, implying that the Coulomb effect plays a small role in these differences.

With the observations in Fig. 2 and Fig. 3, we arrive at the first conclusion of the paper that
imperfect alignment  can play an important role in the PMDs of aligned molecules measured in experiments, and
to study dynamical properties of ATI of molecules aligned at a specific angle in experiments, the effects of imperfect alignment need to be minimized.
The influence of imperfect alignment on PMDs arises from the fact that ionization yields of molecules can differ remarkably at different alignment angles.
For example, for N$_2$, the yields at $\theta=0^0$ are one order of magnitude higher than those at $\theta=90^0$.
As a result, the main contributions to PMD in experiments with imperfect alignment may not come from the aligned sample.
The alignment dependence of ionization of N$_2$ has been discussed in detail in \cite{ADK1} with the MO-ADK theory.
It can also be understood with the interference mechanism under the barrier in the frame of SFA, as discussed in Sec. II. D.

Next, we explore the potential mechanism for the alignment-dependent phenomena in Fig. 3.
In Fig. 4, we plot the offset angles of PMDs of N$_2$ at different angles $\theta$ obtained with different methods.
The TDSE result (the down-triangle curve) shows that the offset angle decreases with the increase of the alignment angle.
The experimental result (the star symbol) agrees with the TDSE prediction. When the MSFA prediction with the full dipole $\textbf{d}_{{{3\sigma_{g}}}}(\textbf{k}_{s})$ of Eq. (10)
(the up-triangle curve) also agrees with the TDSE one,
the MSFA simulations with considering  only the $\sin$-type interference term $\text{sin}({{\textbf{k}_{s}\cdot\textbf{R}}/{2}})$ of Eq. (13) (the square curve)
or only the atomic dipole $\textbf{d}_{2p_z}(\textbf{k}_{s})$ of Eq. (11) (the circle curve) predict the similar tendency
for the alignment-dependent offset angle, with showing a slower decrease of the offset angle for the increase of $\theta$.
When the result of MSFA with only the interference term $\text{sin}({{\textbf{k}_{s}\cdot\textbf{R}}/{2}})$ is somewhat nearer to the full MSFA simulation with the molecular dipole
$\textbf{d}_{{{3\sigma_{g}}}}(\textbf{k}_{s})$ than that with only the atomic dipole $\textbf{d}_{2p_z}(\textbf{k}_{s})$,
we expect that the interference in tunneling associated with the internuclear distance $R$ plays a more important role in this decrease than the contribution of the atomic component $\textbf{d}_{2p_z}$.

\begin{figure}[t]
\begin{center}
\rotatebox{0}{\resizebox *{7cm}{5cm} {\includegraphics {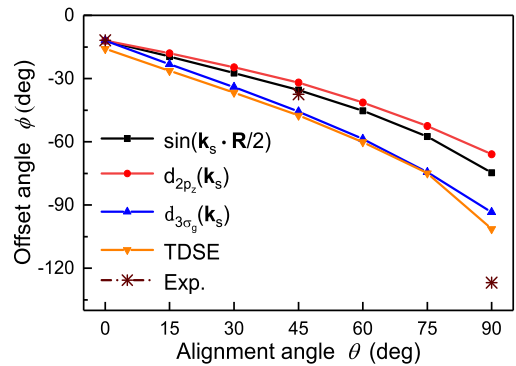}}}
\end{center}
\caption{
The offset angle $\phi$ of PMD of  N$_2$ as a function of the alignment angle $\theta$, obtained through MSFA with
only considering the interference term $\text{sin}({{\textbf{k}_{s}\cdot\textbf{R}}/{2}})$ of Eq. (13) (square),
MSFA with only considering the atomic dipole ${\textbf{d}_{2p_{z}}}(\textbf{k}_{s})$ of Eq. (11) (circle),
MSFA with considering the full dipole $\textbf{d}_{{{3\sigma_{g}}}}(\textbf{k}_{s})$ of Eq. (10) (up-triangle),
TDSE (down-triangle) and experiments (star). Experimental results are abstracted from distributions in the first row of Fig. 3.
The definition of the offset angle is as shown in Fig. 3. For an anticlockwise (clockwise) rotation of the PMD, we define the offset angle is minus (plus).
}
\label{fig:4}
\end{figure}

With the above analyses, the angle dependence of PMD of N$_2$ can be understood with SFA.
In Eq. (5),
the main contributions to the amplitude $F(\mathbf{p},t_0)$ come from the interference term Eq. (13) and the real part of
the term $e^{iS}$ with $S=a+ib$. That is
\begin{equation}
F(\mathbf{p},t_0)\sim \sin(\frac{i\kappa R \cos\theta_I}{2}) e^{-b}.
\end{equation}
As the term $\theta_I$ is functions of both $t_0$ and $\theta$, the term $b$ is the function of only $t_0$. Then we have
\begin{equation}
F(\mathbf{p},t_0)\sim \sin\big[{i\gamma\cos\theta_I(t_0,\theta)}\big] e^{-b(t_0)},
\end{equation}
with $\gamma=\kappa R/2$. On the whole, the term $F_a=e^{-b(t_0)}$ has the maximal value when the time $t_0$ is near to the peak time of the main component $E_z(t)$ of the elliptical laser field.
The term $F_b=|\sin\big[{i\gamma\cos\theta_I(t_0,\theta)}\big]|$ arrives at the maximal value when the angle $\theta_I(t_0,\theta)$ is near to zero, and it has the minimal value of zero at
$\theta_I(t_0,\theta)=90^o$. The interplay of these two terms $F_a$ and $F_b$ dominates the maximal value of the photoelectron amplitude $|F(\mathbf{p},t_0)|$.
For $\theta=0^o$, the values of these two terms peak at the same $t_0$ corresponding to the drift momentum $p_z=0$. However, for $\theta=45^o$, they peak at different $t_0$. As a result,
the maximal value of $|F(\mathbf{p},t_0)|$ peaks at a certain value $t_0$ which is deviated from the peak time of $E_z(t)$ and
is corresponding to the drift momentum with $p_z\neq0$. Accordingly, the absolute value of the offset angle is larger for $\theta=45^o$ than for $\theta=0^o$.
Note, the Coulomb effect induces the rotation of the whole PMD, relative to the description of SFA where the Coulomb effect is neglected, for both cases of these two alignment angles and
is not the main reason for the difference of the offset angles between them, as shown in Fig. 3.
Generally, due to the interference effect, the maximum of the amplitude $|F(\mathbf{p},t_0)|$ for aligned N$_2$ prefers a trajectory ($\textbf{p},t_0$) at which the angle $\theta_I$ is nearer to zero.
For the reason, for $\theta=90^o$, it is the minor component $E_y(t)$ of the elliptical laser field that dominates the ionization of N$_2$.
As a result, the case of $\theta=90^o$ shows a larger absolute offset angle.
We mention that in Fig. 4, the absolute value of the experimental offset angle at $\theta=90^o$  is somewhat larger than the theoretical one.
The reason may be that the ionization signal of the system at the specific angle of $\theta=90^o$ is weak and therefore
the influence of imperfect alignment can not be fully eliminated in obtaining Fig. 3(a3).

\begin{figure}[t]
\begin{center}
\rotatebox{0}{\resizebox *{8.5cm}{6.5cm} {\includegraphics {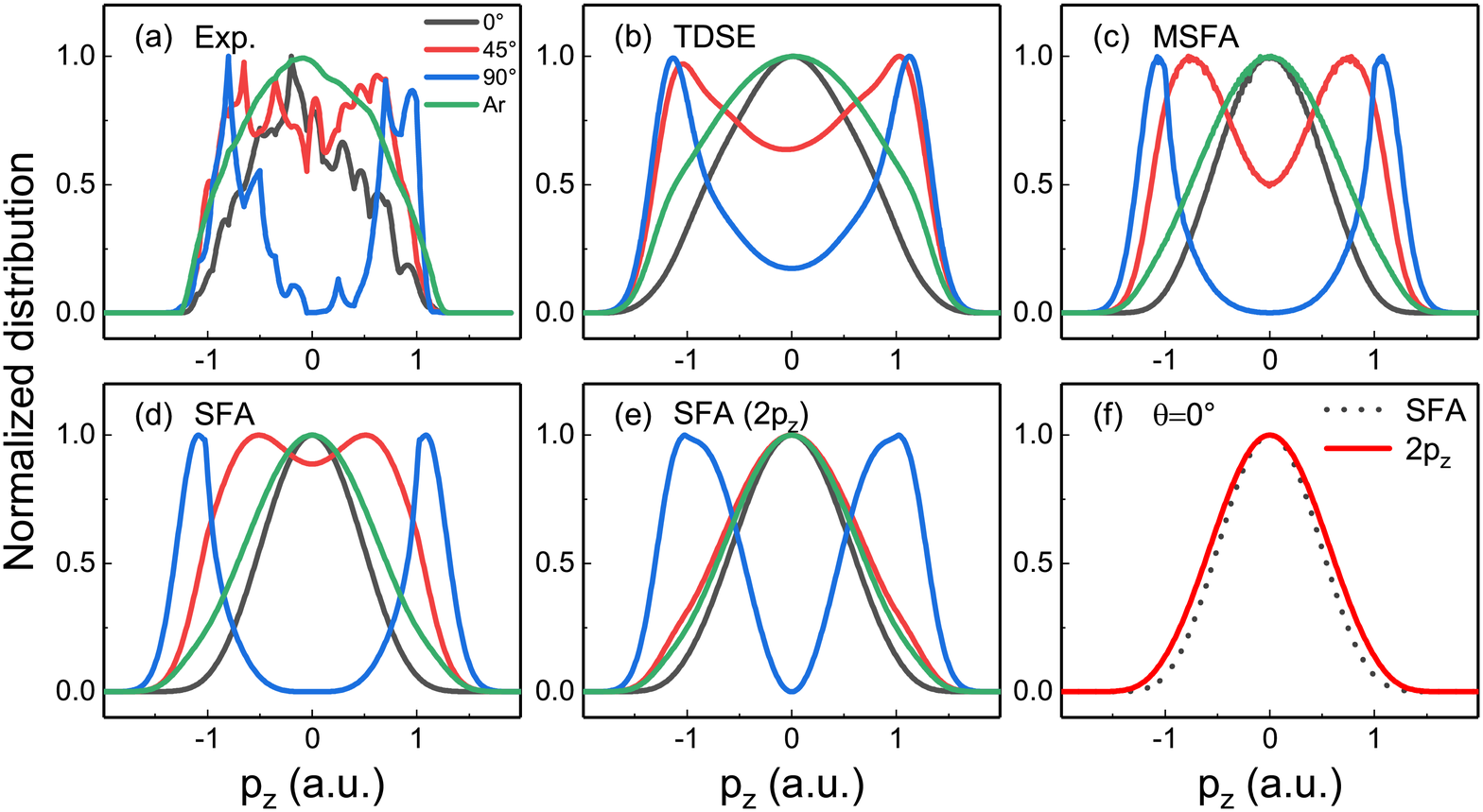}}}
\end{center}
\caption{
Distributions of transverse momentum of N$_2$ at different angles $\theta$ and Ar, obtained with experiments (a), TDSE (b), MSFA (c),
SFA with considering the full dipole $\textbf{d}_{{{3\sigma_{g}}}}(\textbf{k}_{s})$ of Eq. (10) (d),
SFA with only considering the atomic dipole ${\textbf{d}_{2p_{z}}}(\textbf{k}_{s})$ of Eq. (11)  (e).
In (f), we show the direct comparison between the results of $\theta=0^o$ in (d) and (e).
Experimental results for aligned N$_2$  are abstracted from distributions in the first row of Fig. 3 and those for Ar are from Fig. 2(d).
TDSE and model results for Ar are calculated with $1s$ initial state to mimic the randomly-distributed Ar in experiments.
All of the results are normalized to the maximal amplitude of the corresponding distribution.
}
\label{fig:4}
\end{figure}

Besides the offset angle, another characteristic quantity which can be used to quantitatively study the interference effect on PMD is the distribution of transverse momentum, as shown in Fig. 5.
Here, we plot relevant distributions for N$_2$ at these three typical angles and for the reference atom, obtained with different methods.
First, the results of experiments, TDSE and MSFA, shown in the first row of Fig. 5, are similar. When the distribution at $\theta=0^o$ shows a single-peak structure
with the position of the peak around $p_z=0$,  the distributions at $\theta=45^o$ and $90^o$ show the two-peaks one with these two peaks appearing around $p_z=\pm 1$.
The distribution for atom is similar to but somewhat wider than that of $\theta=0^o$. These main characteristics for angle dependence of transverse-momentum distribution also
appear in Fig. 5(d) of SFA simulations, implying that the Coulomb effect plays a small role here, in agreement with our previous discussions.
For SFA simulations with  only considering the atomic dipole of Eq. (11), the distributions at $\theta=0^o$ and $\theta=45^o$ are near to the atomic one, as shown in Fig. 5(e).
This phenomenon is in remarkable disagreement with the SFA prediction of Eq. (10) in Fig. 5(d),
indicating  the important contributions of the interference term $\text{sin}({{\textbf{k}_{s}\cdot\textbf{R}}/{2}})$ included in Eq. (10) to the results in Fig. 5(d).

To highlight the interference effect of Eq. (13) on the distribution of transverse momentum,
in Fig. 5(f), we give a direct comparison of the results at $\theta=0^o$ in Fig. 5(d) with the interference term and Fig. 5(e) without the interference term.
One can observe that the distribution with the interference term is narrower than that obtained with neglecting the interference term in calculations.
This phenomenon suggests that  the interference effect under the barrier affects remarkably on
the distribution of transverse momentum of the aligned molecule.
This is easy to understand. As discussed below Eq. (15), the interference effect will remarkably increase the amplitude of the brightest part of the PMD of the molecule
at $\theta=0^o$, making the whole distribution of transverse momentum of the molecule sharper in comparison with the result without considering the interference effect.
As a result, the relative width of the transverse-momentum distribution for the molecule becomes narrower.

The above results support our another main conclusion in the paper that the interference effect under the barrier plays an important role in alignment dependence of PMD of N$_2$ in elliptical laser fields.

\section{Conclusions}
In summary, we have studied the alignment dependence of ionization of N$_2$ in strong elliptical laser fields experimentally and theoretically.
Due to that the ionization yields of N$_2$ differ remarkably for different alignment angles and perfect alignment is impossible in experiments,
photoelectron momentum distributions (PMDs) of aligned N$_2$ measured in experiments
can include a great deal of information of imperfect alignment, which will preclude a clear identification of alignment dependence of ATI of the molecule in elliptical laser fields.
With the synchronous measurement of aligned N$_2$ and Ar, we are able to highlight the alignment effect on ATI and minimize the effect of imperfect alignment.
When the measured PMDs of N$_2$ are not sensitive to the molecular alignment,
the differences between PMDs of aligned N$_2$  and Ar depend strongly on the alignment angle $\theta$ (the angle between the molecular axis and the main axis of the laser ellipse).
This strong dependence is in good agreement with predictions of TDSE and MSFA (a Coulomb-modified semiclassical SFA model) where perfect alignment is assumed.

In particular, SFA simulations without considering the Coulomb effect also give a good description of this strong dependence,
suggesting that this strong dependence arises mainly from quantum effects associated with the molecular properties in strong laser fields.
With performing analyses based on SFA, we show that the interplay of two-center interference (which is related to the molecular bond length)
and tunneling (which is related to the electron motion under the laser-Coulomb-formed barrier) plays an important role
in the alignment dependence of PMD of N$_2$ in elliptical laser fields.
Due to this interplay, the offset angle of PMD (which is closely related to the most-probable emission angle of the photoelectron) changes remarkably with the alignment angle.
For example, for $\theta=0^0$, the photoelectron of N$_2$ prefers to escape along the minor axis of the laser ellipse, while for
$\theta=90^0$, the main axis of the laser ellipse is preferred.
When the offset angle of PMD is often used as a characteristic quantity to deduce the temporal information of the dynamical process of the target in attosecond measurement,
our results show that the molecular properties already enter the offset angle and  need to be fully considered in future experimental study of
attosecond probing of aligned molecules based on photoelectron spectra.

Besides the offset angle, this interplay also influences remarkably on the distribution of transverse momentum of aligned N$_2$.
For the typical case of $\theta=0^o$, when the interference effect under the barrier remarkably increases the ionization yields of the molecular system
in comparison with the reference atom with similar $I_p$ and orbital components to N$_2$,
it also induces the whole distribution of transverse momentum of the molecule shaper than the atomic one.
As a result, the width of this distribution for the molecule is narrower than the reference atom.

Our work suggests a manner for studying the alignment effect on strong-field ionization of molecules, which is important in ultrafast measurement and control of electron motion inside molecules.

\section{Acknowledgments}

This work was supported by the National Basic Research Program of China (Grant No. 2019YFA0307701),
the National Natural Science Foundation of China (Grants No. 12074143, No. 11627807, No. 12004133, No. 91750111)
and the National Key Research and Development Program of China (Grant No. 2018YFB0504400).


\end{document}